\documentstyle[twoside,psfig]{elsart}
\begin{document}
\begin{frontmatter}
\title	{Fractional calculus and continuous-time finance II: \\
  the waiting-time distribution}

\author[Bologna] {Francesco Mainardi\thanksref{mail*}},
\author[Genova] {Marco Raberto},
\author[Berlin] {Rudolf Gorenflo},
\author[Alessandria,Torino]{Enrico Scalas}
\address[Bologna] {Dipartimento di Fisica, Universit\`a di Bologna and
		  INFN Sezione di Bologna, \\
		  via Irnerio 46, I--40126  Bologna, Italy}
\address[Genova]  {Dipartimento di Ingegneria Biofisica ed Elettronica,
			Universit\`a di Genova, \\
		via dell'Opera Pia 11a,  I--16145 Genova, Italy}
\address[Berlin]{Erstes Mathematisches Institut, Freie Universit\"at
  Berlin, \\
 Arnimallee  3, D-14195 Berlin, Germany}
 \address[Alessandria]	 {Dipartimento di Scienze e Tecnologie Avanzate,
			Universit\`a del \\ Piemonte Orientale,
			via Cavour 84,	I--15100 Alessandria, Italy} 
\address[Torino] {INFN Sezione di Torino, via P.Giuria 1,
			I--10125 Torino, Italy}
\thanks[mail*]{Corresponding author. Fax: +39-051-247244 \\
 {\it E-mail address}: {\tt mainardi@bo.infn.it} (F. Mainardi)}

\begin{abstract}
We complement the theory of
tick-by-tick dynamics  of financial  markets
based on a continuous-time random walk (CTRW) model
recently proposed by Scalas et al \cite{ScaGorMai 00},
and we point out its consistency with
the  behaviour observed in the waiting-time distribution
for BUND future prices traded at LIFFE, London.

\end{abstract}
\begin{keyword}
{{\it PACS: \ }} 02.50.-r, 02.50.Ey, 02.50.Wp, 89.90.+n \\
Stochastic processes; Continuous-time random walk; Fractional calculus;
 Statistical finance;  Econophysics


\end{keyword}

\end{frontmatter}

\def\eg{{e.g.}\ } \def\ie{{i.e.}\ }  
\def\sg{\hbox{sign}\,}
\def\sgn{\hbox{sign}\,}
\def\sign{\hbox{sign}\,}
\def\e{\hbox{e}}
\def\exp{\hbox{exp}}
\def\ds{\displaystyle}
\def\dis{\displaystyle}
\def\q{\quad}	 \def\qq{\qquad}
\def\lan{\langle}\def\ran{\rangle}
\def\l{\left} \def\r{\right}
\def\lra{\Longleftrightarrow}
\def\arg{\hbox{\rm arg}}
\def\d{\partial}
 \def\dr{\partial r}  \def\dt{\partial t}
\def\dx{\partial x}   \def\dy{\partial y}  \def\dz{\partial z}
\def\rec#1{{1\over{#1}}}
\def\log{\hbox{\rm log}\,}
\def\erf{\hbox{\rm erf}\,}     \def\erfc{\hbox{\rm erfc}\,}
\def\FT{\hbox{{F}}\,}  \def\F{\hbox{{F}}\,}  
\def\LT{\hbox{{L}}\,}	\def\L{\hbox{{L}}\,} 
\def\NN{\hbox{\bf N}}
\def\RR{\hbox{\bf R}}
\def\CC{\hbox{\bf C}}
\def\ZZ{\hbox{\bf Z}}
\def\II{\hbox{\bf I}}


\section{Introduction}

In financial markets, not only prices can be modelled as random variables,
but also waiting times between two consecutive
transactions vary in a stochastic fashion.
This fact is well known in financial research. In
his 1973 paper \cite{Clark 73}, Peter Clark wrote:
``Instead of indexing [time] by the integers 0,1,2,$\ldots$, the [price]
process could be indexed by a set of numbers $t_{1}$, $t_{2}$, $t_{3}$,
$\ldots$, where these numbers are themselves a realization
of a stochastic process (with positive increments, so that
$t_{1} < t_{2} <t_{3} < \ldots $).''

Till today, there have been various studies
on the nature of the  stochastic process generating
the sequence of the $t_j$.
In Clark's approach, the variable $t$ is not a physical time
but  an economic variable, the trading volume, an observable whose
increments represent the market intensity.

Lefol and Mercier have written
a review \cite{LefolMercier 98} on the works inspired by Clark's seminal
paper.	In a review by Cont \cite{RamaCont 99},
readers can find pointers to the
relevant literature and a description of the main research
trends in this field.

In a recent paper, Scalas et al. \cite{ScaGorMai 00}
have argued that the continuous time random walk (CTRW) model,
formerly introduced in Statistical Mechanics  by Montroll and
Weiss \cite{Montroll 65}
(on which the reader can find further information in Refs.
\cite{Montroll 79,WeissRubin 83,Montroll 84,Weiss 94b,
Hughes 95,Shlesinger 96,Balescu 97}),  
can provide a phenomenological description of tick-by-tick dynamics
in financial markets. Here, we give further theoretical arguments
and test the theoretical predictions on the waiting-time distribution
against empirical data.

The paper is divided as follows. Section 2 is devoted to the discussion of
a new form for the general master equation in the case of non-local and
non-Markovian processes. In Section 3, the conditions for
the derivation of the time-fractional master equation are given. The
Mittag-Leffler function plays a central role in this respect.
In Section 4, the theoretical predictions on the waiting-time distribution
are compared to market data: high-frequency BUND future prices traded at
LIFFE\footnote{LIFFE stands for London International Financial Futures
(and Options) Exchange. For further information,
 see \	{\tt http://www.liffe.com}.}
in 1997.
Finally, the main conclusions are drawn in Section 5.

\section{The general master equation and the "memory function"}

Throughout this paper the variable $x$ represents the log-price. In other
words, if $S$ is the price of an asset, $x=\log S$. The reason for this
choice is explained by Scalas et al. \cite{ScaGorMai 00}; it is
essentially due to the fact that, rather than prices, returns are the
relevant variable in finance.
The physicist will recognize in $x$ the position of a  random walker
jumping in one dimension. In the following, we shall often use
the random walk language.

Let us consider
the time series $\{ x(t_i) \}\,,$
$\, i=1,2,\dots,$  which
is characterised by $\varphi(\xi, \tau)$, the
{\em joint probability density}
of jumps $\xi_{i} = x(t_{i}) - x(t_{i-1})$ and of waiting times
$\tau_i = t_{i} - t_{i-1}$. The joint density
satisfies the normalization condition
$\int_0^\infty \l[\int_{-\infty}^{+\infty}
 \varphi (\xi, \tau) \,d \xi \r] \,d \tau= 1\,. $
Relevant quantities are the two probability density functions
($pdf$'s)  defined as
$ \, \lambda (\xi) :=  \int_0^\infty
 \varphi (\xi, \tau)  \,d \tau\,,$
$\, \psi(\tau ):= \int_{-\infty}^{+\infty}
 \varphi (\xi, \tau) \,d \xi \,, $
and called {\it jump pdf} and {\it waiting-time pdf}, respectively.

The $CTRW$ is generally defined through the requirement that
the $\,\tau _i\,$   are
identically distributed independent (i.i.d.) random variables.
Furthermore, in the following we shall assume that  the {\it jump} $pdf$
$ \lambda(\xi )$   is independent of  the
{\it waiting-time} $pdf$, $\psi(\tau)\,,$
so that the jumps $\xi_i$ (at instants $t_i\,,$ $\, i = 1,2,3,\dots\,$)
are i.i.d. random variables, all having
the same probability density $ \lambda(\xi)\,. $
Then,  we have
the  factorization
$    \varphi(\xi ,  \tau ) = \lambda   (\xi ) \, \psi ( \tau )\,.$
For convenience we set $t_0=0\,. $

The {\it jump} $pdf$  $ \lambda(\xi) $ represents
 the  $pdf$ for transition
of the walker from a point $x$ to a point $x+\xi \,, $
so it is also called
the {\it  transition} $pdf$.
The  {\it waiting-time} $pdf$
represents the $pdf$ that
a step is taken at the instant $t_{i-1} + \tau $ after the previous one
that happened at the   instant $t_{i-1}\,, $
so it is also called
the {\it pausing-time} $pdf$.
Therefore, the probability that $\tau \le t_i - t_{i-1} < \tau +d\tau $
is equal  to $\psi(\tau) \, d\tau \,. $

The probability that a given interstep interval is greater or equal
to $\tau $ will be denoted by $\Psi(\tau)\,, $
which is defined in terms of $\psi(\tau )$ by
  $$ \Psi(\tau) =\int_\tau ^\infty \psi(t')\, dt'
  = 1- \int_0^\tau  \psi(t')\, dt'\,, \q
 \psi(\tau ) = - {d \over d\tau} \Psi(\tau)\,. \eqno(2.1)$$
We note that  $\int_0^\tau  \psi(t')\, dt'\,$ represents the
probability  that at
least one  step is taken at some instant in the interval $[0,\tau) $,
hence	$\Psi(\tau )\,$   is the probability  that the diffusing quantity
$x$ does not change value during the time interval of duration $\tau $
after a jump.
We also note, recalling that $t_0=0\,,$
that $\Psi(t)\,$
is the {\it survival probability}
until time instant $t$ at the initial
position $x_0=0\,. $

Let us now denote by $p(x,t)$	the $pdf$ of finding the random walker
at the position $x$ at time instant $t\,. $  As usual we assume
the initial condition $p(x,0) = \delta (x)\,, $  meaning that the
walker is initially at the origin $x=0\,. $
We look for the evolution equation for $p(x,t)\,, $
that we shall call	  the {\it master equation} of the $CTRW.$
 Montroll and Weiss \cite{Montroll 65} have shown that the
Fourier-Laplace transform of $p(x,t)$  satisfies a
characteristic equation, now called the {\it Montroll-Weiss equation}.
However, only based upon the previous probabilistic arguments
and without detour onto the Fourier-Laplace domain,
 we can write, directly in the space-time   domain,
  the required {\it master equation}, which reads
$$   p(x,t) =  \delta (x)\, \Psi(t) +
   \int_0^t   \psi(t-t') \, \l[
 \int_{-\infty}^{+\infty}  \lambda(x-x')\, p(x',t')\, dx'\r]\,dt'
 \,. \eqno(2.2) $$
The spatially  discrete analogue of this purely integral form of the
 {\it	master equation}  is quoted  in Klafter et al.  \cite{Klafter 87}
(see also    Ref. \cite{Hilfer-Anton 95}).
We  recognize from Eq. (2.2) the role of the {\it survival probability}
$\Psi(t)$ and of the $pdf$'s $\psi(t)\,,\,  \lambda(x)\,.$
The first term in the RHS of (2.2) expresses the persistence
(whose strength decreases with	 increasing time)
of the initial position $x=0$.
The second term (a spatio-temporal convolution) gives
the contribution to $p(x,t)$ from the walker sitting
in point $x' \in \RR$ at instant $t' < t$ jumping to
point $x$ just at instant $t\,,$  after stopping (or waiting) time
$t-t'\,. $

Now, passing to the Fourier-Laplace domain,
we can promptly derive the celebrated
 {\it Montroll-Weiss equation} \cite{Montroll 65}.
In fact, by adopting
the following standard notation for the generic Fourier and Laplace
transforms:
$$ \F\l\{f(x);\kappa\r\} = \widehat f(\kappa )
 = \int_{-\infty}^{+\infty} \e^{\ds \,i\kappa x}\, f(x)\, dx\,, \q
  {\L} \l\{ f(t);s\r\}= \widetilde f(s)
 = \int_0^{\infty} \e^{\ds \, -st}\, f(t)\, dt\,,  $$
we get from (2.2) the {\it Montroll-Weiss equation}:
$$ \widehat{\widetilde p}(\kappa ,s)
  =  \widetilde {\Psi}(s) \,
 {1 \over 1- \widehat \lambda	 (\kappa )\, \widetilde \psi(s)}
 = {1-\tilde\psi(s)  \over s}\,
 {1 \over 1- \widehat  \lambda(\kappa )\,\widetilde \psi(s)}\,.
 \eqno(2.3) $$

Hereafter we present an alternative form to Eq. (2.2) which
involves the first time derivative of $p(x,t)$ (along with
an additional auxiliary function) so
that the resulting equation  can be interpreted as
an {\it evolution} equation
of {\it Fokker-Planck-Kolmogorov} type.

For our purposes
we re-write Eq. (2.3)  as
 $$  \widetilde \Phi(s) \, \l[
s\,\widehat{\widetilde	p}(\kappa ,s)-1\r] \,= \,
   \,\l[\widehat \lambda   (\kappa)-1\r]\,
   \widehat{\widetilde	p}(\kappa ,s) \,,
\eqno(2.4)$$
where
$$   \widetilde{\Phi} (s) = {1- \widetilde{\psi}(s) \over
	s\, \widetilde{\psi}(s)}
= {\widetilde{\Psi}(s) \over \widetilde{\psi}(s)}
 = {\widetilde{\Psi}(s) \over 1-s\widetilde{\Psi}(s)}  \,.
\eqno(2.5)$$
Then  our {master equation} reads
 $$ \int_0^t   \Phi(t-t')\,
 {\d \over \d t'} p(x,t')\, dt' \, =
     \, -p(x,t) + \int_{-\infty}^{+\infty} \lambda (x-x')\,p(x',t)\,dx'
      \,, \eqno(2.6) $$
where the "auxiliary" function
$\Phi(t)\,,$ being defined through its Laplace transform in Eq. (2.5),
is such that
$\Psi(t) = \int_0^t\Phi (t-t') \, \psi(t' )\,dt' \,.$
We remind the reader that Eq. (2.6),
combined with the initial condition
$p(x,0) = \delta (x)\,,$ is equivalent to Eq. (2.4),
and then
its solution represents the {Green
function} or the {fundamental solution} of the	
Cauchy problem.

From Eq. (2.6) we recognize the role of "memory function"    for
$\Phi(t)\,.$
As a  consequence, the CTRW turns out to be
in general a non-Markovian process.
However, the process is "memoryless",
namely "Markovian"  if (and only if)  the above
memory function degenerates into a delta function (multiplied
by a certain positive constant) so that $\Psi(t)$ and $\psi(t)$ may
differ only by a multiplying positive constant.
By appropriate choice of the unit of time
 we assume
$  \widetilde \Phi(s) = 1\,,$ so
 $ \Phi(t) =  \delta (t)\,,\; t\ge 0\,.$
In this case we  derive
$$
\widetilde\psi(s)  =\widetilde \Psi(s)
 = {1\over 1+s} \,,\q \hbox{so} \q
 \psi(t) =  \Psi(t)= \e^{\,\ds -t}\,,\; t\ge0\,.
 \eqno(2.7)$$
Then Eq. (2.6) reduces to
$$  {\d \over \d t} p(x,t) = - p(x,t)  +
   \int_{-\infty}^{+\infty}   \lambda	(x-x')\,p(x',t)\,dx'\,,
     \q p(x,0) = \delta(x)\,. \eqno(2.8) $$
 This is, up to a change of the unit of time
 (which means multiplication of the R.H.S by a positive constant),
 the most general {\it master equation} for a {\it Markovian}
 $CTRW$; it is called	the {\it Kolmogorov-Feller equation}
in Ref. \cite{Saichev 97}.

We note that the form (2.6), by exhibiting  a weighted first-time
derivative,  is original as far as we know;
it allows us to characterize in a natural way	a peculiar
class of non-Markovian processes, as shown in the next Section.

\section{The time-fractional master equation for "long-memory"
processes}

Let us now consider "long-memory" processes, namely
non-Markovian processes characterized by a memory function
$\Phi(t)\,$
exhibiting a power-law	time decay.
To this purpose a natural choice is
$$
    \Phi(t)   = {t^{-\beta}\over  \Gamma(1-\beta)}\,,  \q t\ge 0\,,
 \q 0<\beta <1\,. \eqno(3.1)$$
Thus,  $\Phi(t)$ is a weakly singular function
that, in the limiting case $\beta =1\,, $
reduces to  $\Phi(t) =	\delta (t)\,, $
according to  the formal representation of the Dirac generalized function,
$\delta(t) = t^{-1}/\Gamma(0)\,, \; t \ge 0\,$
(see \eg REf. \cite{Gel'fand 64}).

As a consequence of the choice Eq. (3.1), we  recognize that
(in this peculiar non-Markovian situation)
our {\it master equation}  (2.6)  contains a time
fractional derivative.
  In fact, by inserting into Eq. (2.4) the Laplace transform of $\Phi(t)\,,$
$\widetilde \Phi(s) = 1/s^{1-\beta }\,, $
 we get
$$ s^\beta  \, \widehat{\widetilde  p}(\kappa ,s) - s^{\beta -1}
    = \l[\widehat \lambda   (\kappa)-1\r]\,
   \widehat{\widetilde	p}(\kappa ,s) \,,
 \q 0<\beta < 1\,,
\eqno(3.2) $$
so  that the resulting Eq.  (2.6) can be written as
$$ {\d^\beta   \over \d t^\beta } p(x,t) =
     -	p(x,t) +   \int_{-\infty}^{+\infty} \lambda(x-x')\,
   p(x',t) \, dx'\,, \q p(x,0) = \delta(x)\,,
\eqno(3.3) $$
where ${\d^\beta  / \d t^\beta } $
is the pseudo-differential operator
explicitly defined in the Appendix, that we call the
{\it Caputo} fractional derivative of order $\beta  \,. $
Thus Eq. (3.3) can be considered
as the {time-fractional
generalization} of Eq. (2.8) and consequently can be called
the {\it time-fractional \  Kolmogorov-Feller equation}.
We note that this derivation differs from the one
presented in Ref. \cite{ScaGorMai 00}
and references therein,
in that here we have pointed out the
role of the long-memory processes rather than that of scaling
behaviour in the hydrodynamic limit. Furthermore
here the {\it Caputo} fractional derivative appears in a natural way
without use of the
{\it Riemann-Liouville} fractional derivative.

Our choice  for $\Phi(t)$  implies peculiar forms
for the functions $\Psi(t)$ and $\psi(t)$  that 
generalize the exponential behaviour (2.7) of the Markovian case.
In fact, working in the Laplace domain we get from (2.5) and (3.1)
 $$ \widetilde \Psi(s) =
 {s^{\beta-1} \over 1+ s^\beta}\,,   \q
 \widetilde \psi(s)= {1 \over 1+  s^{\beta}}\,,\q
    0<\beta < 1\,,\eqno(3.4) $$
from which by inversion we obtain for $t\ge 0$
$$\Psi(t) =  E_\beta (-t^\beta)
\,,\q \psi(t) =  
	    -	{d \over dt}  E_\beta (-t^\beta)
 \,, \q 0<\beta < 1\,,\eqno (3.5)$$
where $E_{\beta}$ denotes an entire transcendental function, known as
the Mittag-Leffler function of order $\beta\,,$
defined in the complex plane by the power series
$$ E_\beta (z) :=
    \sum_{n=0}^{\infty}\,
   {z^{n}\over\Gamma(\beta\,n+1)}\,, \q \beta >0\,, \q z \in \CC\,.
 \eqno	(3.6)$$
For detailed information on the Mittag-Leffler-type functions
and their Laplace transforms the reader  may  consult \eg
\cite{Erdelyi HTF,GorMai 96,GorMai 97,MaiGor 00}.
We note that for $0<\beta <1$ and $1<\beta <2$	the function
$\Psi(t)$  appears
in certain  relaxation and oscillation processes,
then called  {\it fractional relaxation} and {\it fractional oscillation}
processes, respectively
(see \eg  Refs. 
 \cite{GorMai 96,GorMai 97,Mainardi 96a,Mainardi 97}
 and references therein).

Hereafter, we find it convenient to summarize
the  features of the functions $\Psi(t)$ and $\psi(t)$	 most relevant for
our purposes. We begin to quote their series expansions
and  asymptotic representations:
$$ \Psi(t) \,
  \cases{
  \,= \,{\ds \sum_{n=0}^{\infty}}\,
  (-1)^n {\ds {t^{\beta n}\over\Gamma(\beta\,n+1)}}\,,
      & $\, t\ge 0$\cr\cr
  \, \sim \,  {\ds {\sin \,(\beta \pi)\over \pi}}
  \,{\ds  {\Gamma(\beta)\over t^\beta}}\,,
  & $\, t\to \infty \,,$}
     \eqno(3.7) $$
and
$$ \psi(t) \,
  \cases{
  \,= \,{\ds {1\over t^{1-\beta}}}\, {\ds \sum_{n=0}^{\infty}}\,
  (-1)^n {\ds {t^{\beta n}\over\Gamma(\beta\,n+\beta )}}\,,
      & $\, t\ge 0$\cr\cr
  \, \sim \,  {\ds {\sin \,(\beta \pi)\over \pi}}
  \,{\ds  {\Gamma(\beta+1)\over t^{\beta+1}}}\,,
  & $\, t\to \infty \,.$}
     \eqno(3.8) $$
The  expression for $\psi(t)$ can be shown to be equivalent to that one
obtained in  Ref. \cite{Hilfer-Anton 95} in terms of the
generalized Mittag-Leffler function in two parameters.

In the limit for $\beta \to 1$ we recover the exponential functions
of the Markovian case.
We note that for $0<\beta <1$ both  functions
 $\psi(t)$, $\Psi(t)$, even if losing their exponential
decay by exhibiting power-law tails
for large times,  keep	 the "completely monotonic" character.
Complete monotonicity of  the	functions
 $\psi(t)$, $\Psi(t)$, $t>0$, means:
$$ (-1)^n {d^n\over dt^n}\, \Psi  (t) \ge 0\,,	\q
   (-1)^n {d^n\over dt^n}\, \psi  (t) \ge 0\,,
\q n=0,1,2,\dots	\eqno(3.9)$$
or equivalently, their representability as (real) Laplace transforms
of non-negative  
functions.
In fact it can be shown
for $0<\beta <1\,:$
$$ \Psi(t)  =
  {\ds{\sin \,(\beta \pi)\over \pi}\,
   \int_0^\infty \!
   { r^{\beta  -1}\, \e^{\,\ds -rt}\over
    r^{2\beta } + 2\, r^{\beta } \, \cos(\beta	\pi) +1}\, dr}\,,
  \q t \ge 0\,,
\eqno(3.10)$$
 and
$$  \psi(t) =
  {\ds {\sin \,(\beta \pi)\over \pi}\,
   \int_0^\infty \!
   { r^{\beta}\, \e^{\,\ds -rt} \over
    r^{2\beta } + 2\, r^{\beta}\,\cos(\beta \pi) +1}\, dr}\,,
 \q t\ge 0  \,.    \eqno(3.11)	  $$

A special case is $\beta = {1\over 2}$ for which it is known that
$$ E_{1/2} (-\sqrt{t}) =
    \e^{\ds \, t}\, \hbox{erfc} (\sqrt{t}) =
 \e^{\, \ds t}\, {2\over \sqrt{\pi}}\,
  \int_{\sqrt {t}}^\infty \e^{\, \ds -u^2}\,du	\,,\q t\ge 0\,,
\eqno(3.12)$$
where $ \, \hbox{erfc}\,$ denotes the
{\it complementary error} function.

It may be instructive to note that for sufficiently small times
$\Psi(t)$    exhibits a behaviour similar to that of a
stretched exponential; in fact	we have
$$  E_\beta  (- t^\beta ) \simeq
 1 - {t^\beta  \over \Gamma(\beta  +1)}\,
   \simeq \,\exp\{ - t^\beta /\Gamma(1+\beta )\}
 \,, \q 0 \le t \ll  1\,.
   \eqno(3.13)$$

Hereafter, we consider	two relevant forms
for the survival probability $\Psi(t)$
(that we shall denote by $f_\beta(t)$ and $g_\beta(t)$
to distinguish them from $e_\beta (t):= E_\beta(-t^\beta )$)
  which,  in exhibiting  a decreasing behaviour with a
power law decay for large times,
represent
alternative candidates for long-memory processes.

The simplest function which meets these requirements is expected to be:
$$ f_\beta (t) := {1\over  1+ \Gamma(1-\beta) t^\beta } \,,
 \q t\ge 0\,,
\eqno(3.14)$$
so that
$$ f_\beta (t) \,\sim
  \,\cases{
  \, 1 - {\ds {\pi \beta \over \sin(\pi \beta )}} \,
 {\ds {t^\beta\over  \Gamma(1+\beta )}} \,,
     & $\, t \to 0\,,  $\cr\cr
  {\ds {\sin \,(\beta \pi)\over \pi}}
  \,{\ds  {\Gamma(\beta)\over t^\beta}}\,,
 & $\, t \to \infty\,.$ \cr}  \eqno(3.15)  $$
One can infer from the Eqs (3.7) and (3.15)
that $e_\beta  (t)$ and
$f_\beta (t)$ practically coincide for all $t\ge 0$
if $\beta $ is sufficiently small, say for $0<\beta < 0.25\,. $
For  greater values of $\beta\,,$ that are relevant in our subsequent
empirical analysis, their  difference is expected to be
appreciable in a wide range of time intervals.

Another possible choice of major statistical relevance is
based on the assumption that the {\it waiting-time} $pdf$
$\psi(t)$ may be an extremal,  unilateral, stable distribution with
index $\beta \,.$
In this case the Laplace transforms of
$\psi(t)$ and $\Psi(t)$
read
$$ \widetilde\psi(s)  =
   \exp (-s^\beta )\,,\q
 \widetilde \Psi(s) = {1-\exp(-s^\beta)\over s}\,,
\q 0<\beta <1\,.  \eqno(3.16)  $$
By inversion we obtain for $t\ge 0$
$$\psi(t) =
    {1\over t}\, \phi_{-\beta,0} \l(- {1\over t^\beta}\r)\,,
 \q
    \Psi(t) =	 
    1-	\phi_{-\beta,1} \l(- {1\over t^\beta}\r)\,,
  \q 0<\beta < 1\,,\eqno (3.17)$$
where $\phi _{-\beta,0}\,,$  $\,\phi _{-\beta,1}$
denote entire transcendental functions
(depending on two indices), known as
the Wright functions,
defined in the complex plane by the power series
$$ \phi_{\lambda ,\mu } (z) :=
    \sum_{n=0}^{\infty}\,
   {z^{n}\over n!\Gamma(\lambda \,n+\mu )}\,,
 \q \lambda  >-1\,,\q \mu \in \CC\,, \q z \in
\CC\,.	\eqno  (3.18)$$
For detailed information on the Wright type functions
and their Laplace transforms the reader  may  consult \eg
Refs. \cite{Erdelyi HTF,Mainardi 97,GoLuMa 99}.
We note that for $0<\beta <1$ and $\mu = 0$ or
$\mu  =1-\beta $   the corresponding Wright functions
  appear in the fundamental solutions
of {\it time-fractional diffusion} equations
(see \eg
Refs. \cite{Mainardi 96a,Mainardi 97,GoLuMa 00}
and references therein).

Hereafter, like for the Mittag-Leffler-type functions (3.5),
we quote for the Wright-type functions (3.17)
their series expansions
and asymptotic representations.
For the {\it waiting-time} $pdf$  we have
 (see \eg Ref. \cite{Schneider 86}),
  $$ \psi(t) =
  {\ds {1\over \pi t}}\, {\ds \sum_{n=1}^{\infty}}\,
  (-1)^{n-1} {\ds {\Gamma(\beta\,n+1 ) \over n!}}\,
  {\ds {\sin(\pi \beta n)\over t^{\beta n}}} \,, \q t> 0\,,
\eqno(3.19)$$
and
$$\psi(t)  \, \sim \, A\, {\ds t^{-a}} \, \exp \l(
  \,{\ds - B \, t^{-b}}\r)\,,
  \q t\to 0 \,,\eqno(3.20)$$
where
$$    A = \l[{\beta^{1/(1-\beta )}\over 2\pi (1-\beta)}\r]^{1/2}, \;
      a = {2-\beta \over 2(1-\beta)}\,,\q
      B = (1-\beta )\,\beta ^b\,,\;
      b = {\beta \over 1-\beta }\,.\eqno(3.21)$$

For the {\it survival probability} we obtain
$$  g_\beta (t) =\Psi(t) \, = \,
  {\ds {1\over \pi} }\, {\ds \sum_{n=1}^{\infty}}\,
  (-1)^{n-1} {\ds {\Gamma(\beta\,n ) \over n!}}\,
  {\ds {\sin(\pi \beta n)\over t^{\beta n}}} \,, \q t> 0\,,
\eqno(3.22)$$
and   
  $$   g_\beta (t)=  \Psi(t) \,\sim \,
  1- C\, {\ds t^{c}} \, \exp \l(
  \,{\ds - B \, t^{-b}}\r)\,,
  \q t\to 0 \,,\eqno(3.23)$$
where
$$    C =  \l[
     {1\over 2\pi (1-\beta)\, \beta^{1/(1-\beta )}}\r]^{1/2}\,,\q
      c = {\beta \over 2(1-\beta) } = {b\over 2} \,.
\eqno(3.24)$$

Like for the Mittag-Leffler function,
a special case is again  $\beta ={1\over 2}$ 
for which we obtain the analytical
expressions
$$ \psi(t) = {1\over \sqrt{\pi}}\, t^{-3/2}\,\exp \l(- {1\over 4t}\r)\,,
 \q  \Psi(t) = {\ds \hbox{\erf} \l({1\over 2\sqrt{t}}\r)} \,,\q
 t \ge	0\,. \eqno(3.25)$$
We note that in this particular case the asymptotic representation
(3.20) and (3.21) provides the sum of the series (3.19) 
and henceforth the exact
expression for $\psi(t)$ in Eq. (3.25),
the so-called {\it L\'evy-Smirnov pdf}
(see \eg Ref. \cite{Mantegna 00}).

Hereafter we would like to point out the major differences between
the Mittag-Leffler-type function $e_\beta (t):= E_\beta(-t^\beta)$  and
the Wright type function $g_\beta (t):= 1- \phi_{-\beta,1}(-t^{-\beta})\,, $
that can be inferred by analytical arguments.
The first difference  concerns	the decreasing behaviour before the
onset of the common power law regime:
whereas $e_\beta (t)$
starts at $t=0$ vertically (the derivative is $-\infty$)  and is
completely monotone,
$g_\beta (t)$	 starts at $t=0$ horizontally (the derivative is $0$)
and then  exhibits a change in the concavity from downwards to upwards.
A second difference  concerns
the limit for $\beta \to 1\,;$
whereas  $e_\beta (t)$	tends to the
exponential $\exp (-t)$ (no memory),
$g_\beta (t)$ tends to the box function
$H(t)-H(t-1)$ 
(as directly obtained from the Laplace inversion of Eq. (3.16)
for $\beta=1$).
As a consequence,  the corresponding {\it waiting-time pdf} tends to the
Dirac delta  function $\delta(t-1)\,,$
a peculiar case considered by  Weiss \cite{Weiss 94b}
in his book  in p. 47 as an example of a non-Markovian process.


\section{Empirical analysis}

In order to corroborate the theory presented above, we have
analyzed the waiting-time distribution of BUND futures
traded at LIFFE in 1997.
{BUND} is the German word for bond. Futures are derivative
contracts in which a party agrees to sell and the other party
to buy a fixed amount of an underlying asset at a
given price and at a future delivery date. In this case the underlying
asset is a German Government bond.

We have considered two different delivery dates:
June 1997 and
September 1997. Usually, for a future with a certain maturity,
transactions begin some months before the delivery date. At the
beginning, there are few trades a day, but closer to
the delivery there may be more than $1\,000$ transactions a day.
For each maturity, the total number of transaction is greater
than $160\,000$.

In Figs. 1 and 2 we plot $\Psi (\tau)$ for the June and
September delivery dates, respectively. The
circles refer to market data and represent the probability of
a waiting time greater than the abscissa $\tau$. We have
determined about 600 values of $\Psi(\tau)$
for $\tau$ in the interval between $1\,$ and $50\,000\,$s,
neglecting the intervals of market closure.
The solid line is a two-parameter fit obtained
by using the Mittag-Leffler-type function
$$ \Psi(\tau) = e_\beta (\gamma \tau ) =
 E_{\beta} \l[ -(\gamma \tau)^{\beta} \r]\,, \eqno(4.1) $$
where $\beta$ is the index of the  Mittag-Leffler
function and $\gamma$ is a time-scale factor, depending on the time unit.
For the June delivery date we get an index $\beta=0.96$
and a scale factor $\gamma={1\over 12}\,,$ whereas, for the September
delivery date, we have $\beta=0.95$ and $\gamma={1\over 12}\,.$
The fit in Fig. 1 has a reduced chi square
$\widetilde{\chi}^{2} \simeq 0.26$, whereas the reduced chi square of
the fit in Fig. 2 is $\widetilde{\chi}^{2} \simeq 0.25$.
The chi-square values have been computed
considering all the values of $\Psi\,.$ 

In Figs. 1 and 2,
the dash-dotted line is the stretched exponential
function $\exp \{ -(\gamma \tau)^{\beta})/\Gamma (1+\beta)\}\,$
(see Eq. (3.13)), whereas
the dashed line is the power-law function
$(\gamma \tau)^{-\beta}/\Gamma(1-\beta)\,$ 
(see the second equation in Eq. (3.7)).
The Mittag-Leffler function
interpolates between these two limiting behaviours:
the stretched exponential for small time intervals, 
and the power-law
for large ones.

Even if the two fits seem to work well, some words of caution are
necessary. The Mittag-Leffler-type function $e_\beta (\gamma \tau)$
naturally derives from our
assumption on the "memory function" in the CTRW model;
however, as  previously observed,
it is not the unique possibility compatible with a long-memory
process with a power-law decay.  As a consequence, hereafter,
we shall also consider	for $\Psi(\tau )$ the two alternative functions
discussed in Section 3, namely the rational function
$f_\beta (\gamma \tau)\,$ 
(see	Eqs. (3.14) and (3.15)),
and the Wright function $g_\beta (\gamma \tau)\,$ 
(see Eqs. (3.22)-(3.24)).

In  Figs. 3 and 4, by taking the same data as in Figs. 1
and 2 respectively,  we compare
the functions $e_\beta (\gamma \tau )$ (solid line),
$f_\beta(\gamma \tau )$  (dash-dotted line) and
$g_\beta (\gamma \tau )$ (dashed line).
Whereas in the previous figures  we have adopted a log-log scale to point
out the power-law decay by a straight-line, now  we find it convenient
to use a linear scale for the ordinates to point out the behaviour
of the functions for  small values of $\tau \,. $
From these figures we can infer that the Mittag-Leffler
function  fits the data of the empirical
analysis   much better	than the  other two chosen functions,
thus corroborating  our approach to CTRW
based on the {\it fractional-time derivative}.
However, the Mittag-Leffler fit significantly differs from the
empirical data for small values of $\tau \,. $

\section{Conclusions}

The CTRW is a good phenomenological description of the tick-by-tick
dynamics in a financial market. Indeed, the CTRW can naturally take into
account the pathological time-evolution of financial markets, which is
non-Markovian and/or non-local. From this point of view, by a proper
choice of a (perhaps non-stationary) joint $pdf$ $\varphi(\xi,\tau)$,
one could accurately reproduce the statistical properties
of market evolution. In this respect, the model can be useful for
applications where Monte-Carlo simulations of market settings are needed.

With additional assumptions, the CTRW hypothesis can be tested against
empirical data, thus providing useful information on the restrictions of
the premises. In this paper, we have assumed a particular form for the
time-evolution kernel, leading to a time-fractional Kolmogorov-Feller
equation. In its turn, this implies that $\Psi (\tau)$, the probability
of finding a waiting-time interval greater than $\tau$, is a
Mittag-Leffler function. There is a satisfactory agreement between this
prediction and	the empirical distributions analyzed in
Figs. 1-4,
but not for small  time intervals.

Among the various questions for future research on this topic, two are
particularly relevant in our opinion. The first one concerns the
behaviour of other assets. Prices of liquid stocks should have a
completely different time scale. Indeed, for the futures here
considered, at the beginning of their lifetime several hours passed
between two consecutive trades, a feature which is not likely to be shared
by liquid stocks. The second problem concerns the uniqueness of the
kernel. The Mittag-Leffler kernel yields the elegant time-fractional
Kolmogorov equation, but there might be other possibilities for
interpolating between the small waiting-time and the large waiting-time
behaviour of $\Psi(\tau)$.

Finally, there is an implication for microscopic market models, a realm
where many physicists have started researching. We believe that a
microscopic model should, at least phenomenologically, take into account
that agents in the market decide to sell and buy an asset at randomly
distributed instants. It would be a success to derive the ``right''
waiting-time distribution from first principles, whatever these first
principles will be.


 \vskip 0.8truecm 
\noindent {\bf Appendix. The Caputo fractional derivative}

\vskip 0.5truecm

For the sake of convenience of the reader here we present an
introduction to the {\it Caputo} fractional derivative starting from its
representation in the Laplace  domain and pointing out its difference
with respect to the standard {\it Riemann-Liouville} fractional derivative.
So doing,g we avoid
the subtleties lying in the inversion of fractional integrals.
If $f(t)$ is a (sufficiently well-behaved) function  with Laplace
transform
$  \; {\L} \l\{ f(t);s\r\}= \widetilde f(s)
 = \int_0^{\infty} \e^{\ds \, -st}\, f(t)\, dt\,,
$
we have
$$  {L} \l\{ {d^\beta \over d t^\beta} f(t);s\r\} =
    s^\beta \, \widetilde f(s) - s^{\beta-1}\, f(0^+)\,, \q 0<\beta <1\,,
 \eqno(A.1) $$
if we define
$$ {d^\beta \over d t^\beta} \,f(t) :=
      {1 \over \Gamma(1-\beta )}\,\int_0^t
 {df(\tau )\over d\tau }\, {d\tau \over (t-\tau )^{\beta}} \,.
 \eqno(A.2) $$
We can also write
$$ {d^\beta \over d t^\beta} f(t)=
  {1 \over \Gamma(1-\beta)}\,{d  \over d t} \l\{
   \int_0^t
   [f(\tau )- f(0^+)]\,
  {d \tau \over (t-\tau )^{\beta}} \r\}\,,
\eqno(A.3)$$ 
   $$ {d^\beta \over d t^\beta} f(t)=
  {1 \over \Gamma(1-\beta )}\,{d \over d t} \l\{
    \int_0^t
  {f(\tau )\over (t-\tau )^{\beta}} \,d \tau \r\}
  - {t^{-\beta }\over \Gamma(1-\beta)}\, f(0^+) \,. \eqno(A.4)$$

A modification of Eqs. (A.1)-(A.4) 
holds any non integer $\beta >1\, $
(see \cite{GorMai 97}).
We refer to the fractional derivative defined by Eq. (A.2)	as
the {\it Caputo} fractional derivative, since it was formerly
applied by Caputo in the late 1960s 
for modelling  the dissipation effects in {\it Linear Viscoelasticity}
(see  \eg Refs. \cite{Caputo 67,Mainardi 97}).
The reader should observe that this definition
differs from the usual one named after
Riemann  and Liouville, which is given by the first term in
the R.H.S. of (A.4)
see \eg \cite{Samko 93}.
For more details on the {\it Caputo} fractional derivative
we refer to  Refs. \cite{GorMai 97,Podlubny 99,Butzer 00}.

\vfill\eject


$\null$
\vskip 0.1 truecm
\begin{center}
\begin{figure}[htbp]
 \centerline{\psfig{file=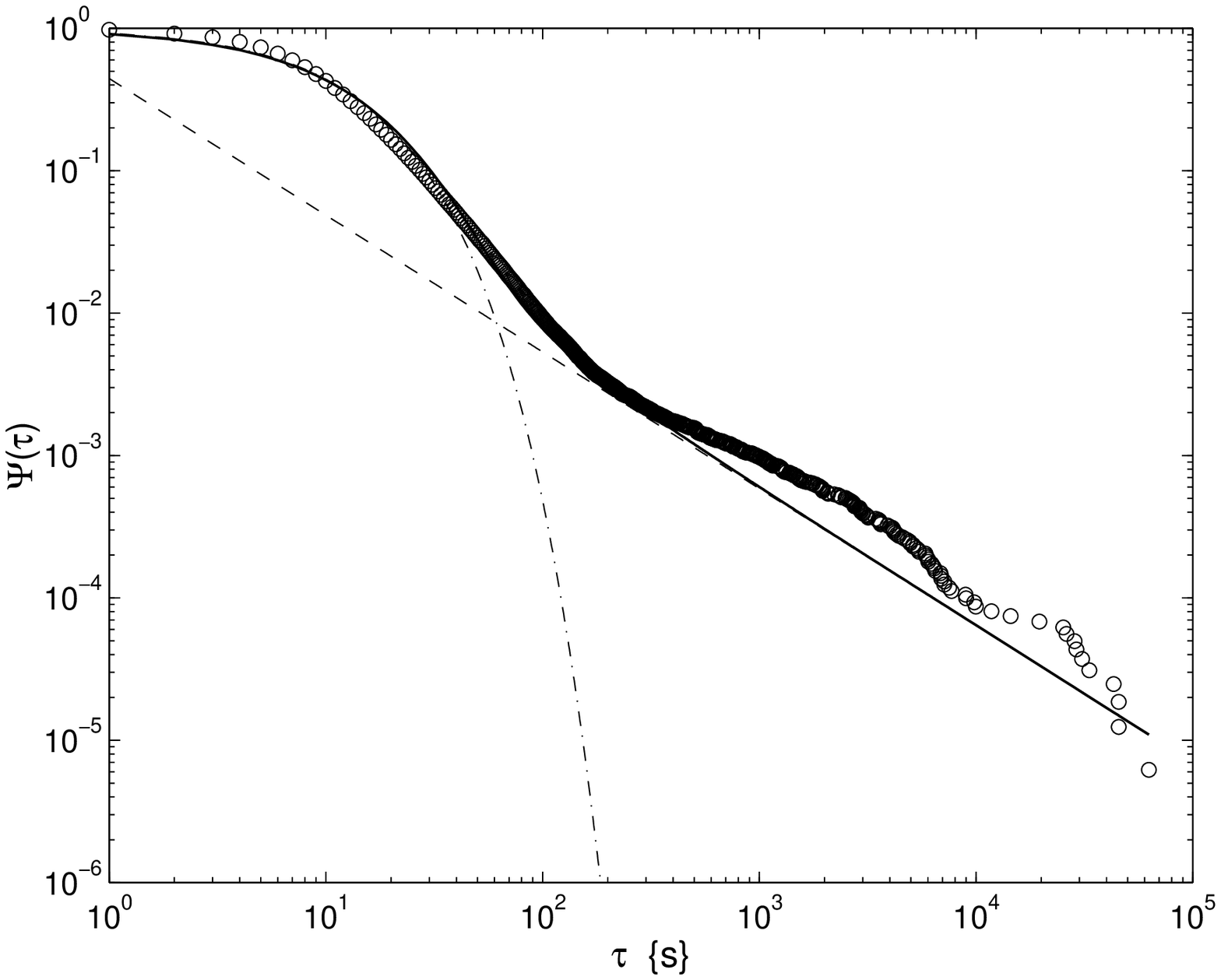,height=7.5truecm,width=10.0truecm}}
\end{figure}
\center{\footnotesize { {\bf Fig. 1}\\
Survival probability for {\it BUND} futures with delivery date:
June 1997.\\
The Mittag-Leffler function (solid line) is compared with the stretched \\
exponential (dash-dotted line) and the power (dashed line) functions. \\
($\beta =0.96\,,\; \gamma =1/12$)}}
 \end{center}

$\null$
\vskip 0.1truecm
\begin{center}
\begin{figure}[htbp]
\centerline{\psfig{file=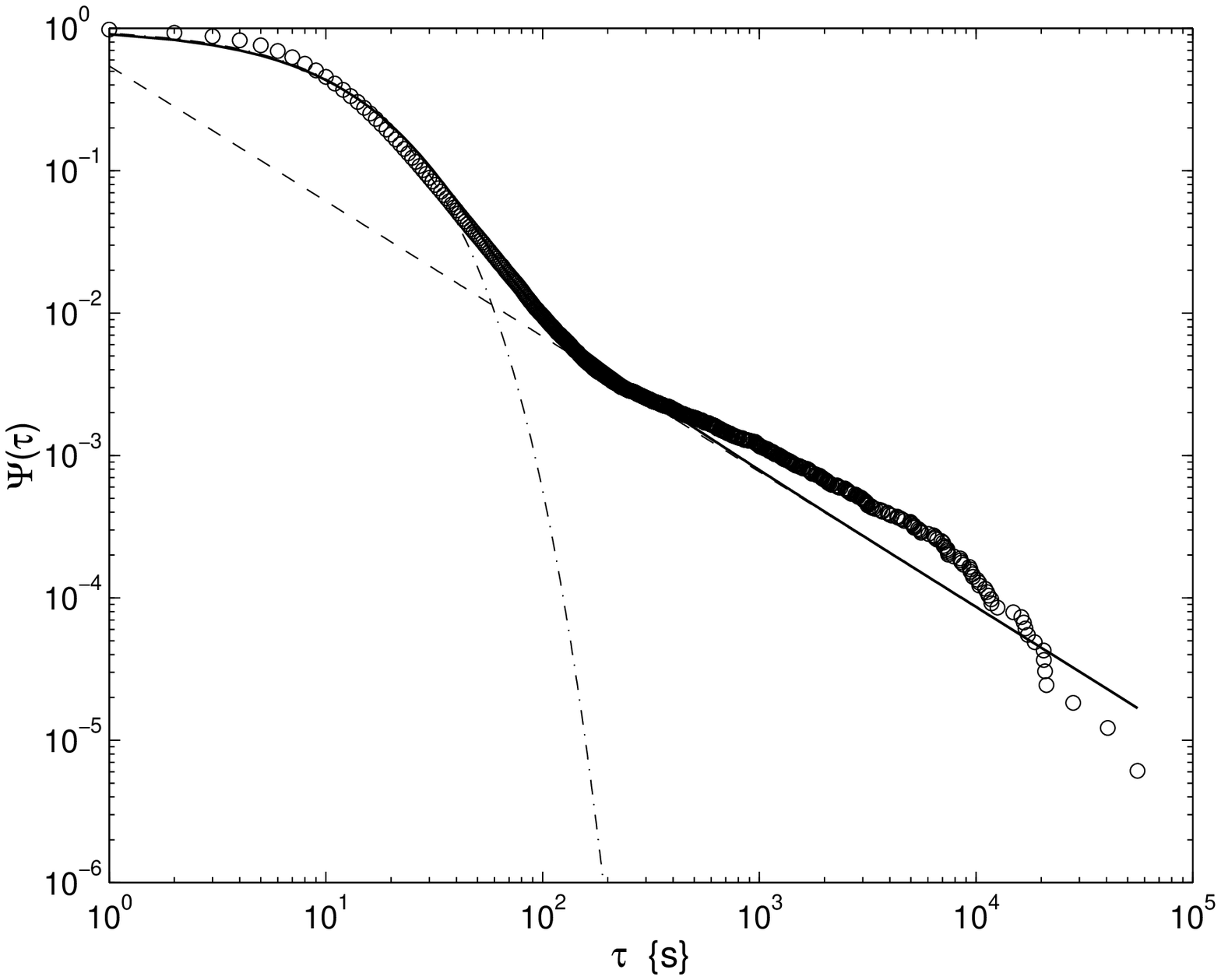,height=7.5truecm,width=10.0 truecm}}
\end{figure}
\center{\footnotesize { {\bf Fig. 2}\\
Survival probability for {\it BUND} futures with delivery
  date: September  1997. \\
The Mittag-Leffler function (solid line) is compared with the stretched \\
exponential (dash-dotted line) and the power (dashed line) functions.\\
($\beta =0.95\,,\; \gamma =1/12 $)}}
 \end{center}
\vfill\eject

$\null$
\vskip 0.1 truecm
\begin{center}
\begin{figure}[htbp]
\centerline{\psfig{file=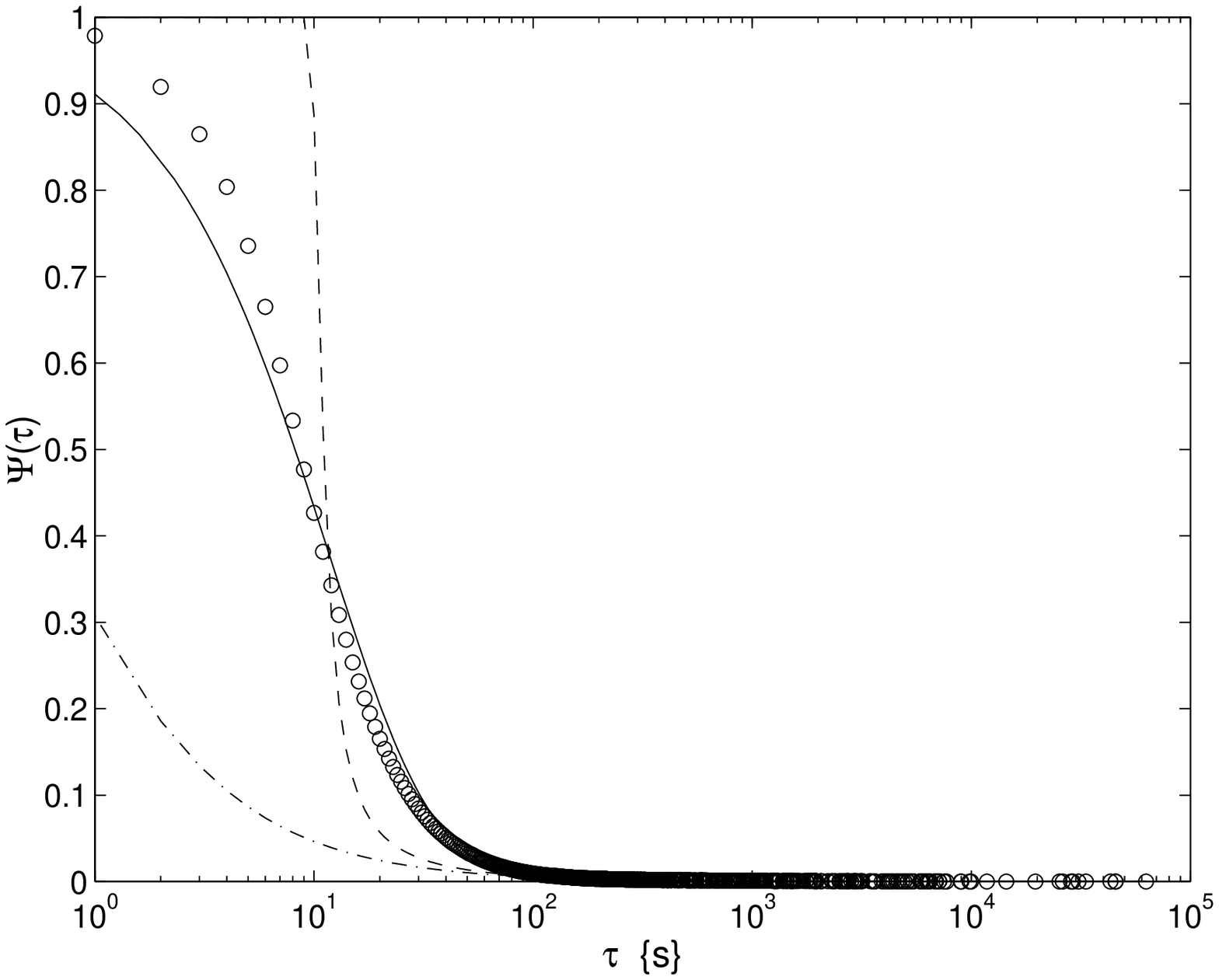,height=7.5truecm,width=10.0truecm}}
\end{figure}
\center{\footnotesize { {\bf Fig. 3}\\
Survival probability for {\it BUND} futures with delivery date:
 June	1997.\\
The Mittag-Leffler function (solid line) is compared with the rational \\
(dash-dotted line) and	the Wright (dashed line) functions.\\
($\beta =0.96\,,\; \gamma =1/12 $)}}
\end{center}


$\null$
\vskip 0.1truecm
\begin{center}
\begin{figure}[htbp]
\centerline{\psfig{file=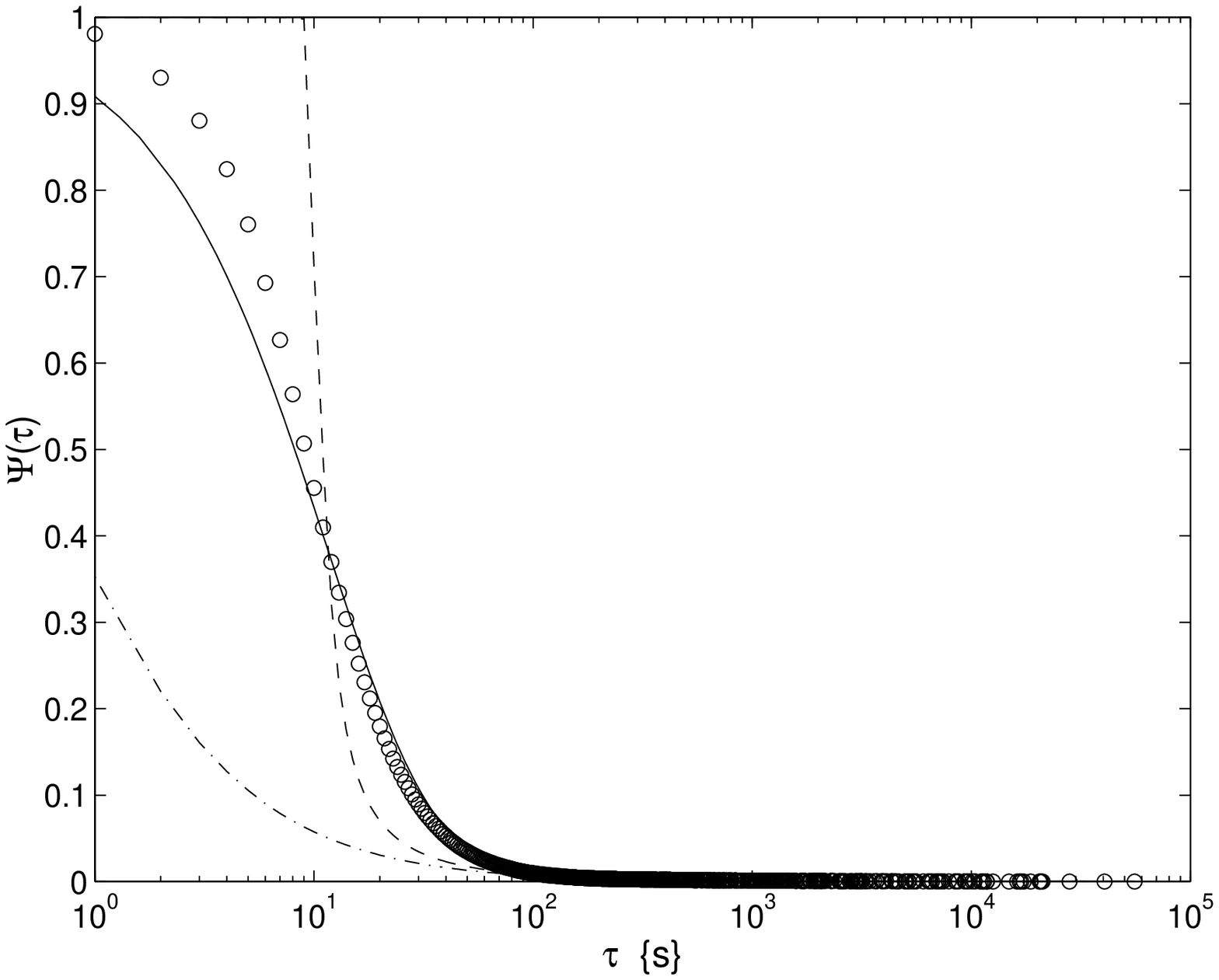,height=7.5truecm,width=10.0truecm}}
\end{figure}
\center{\footnotesize { {\bf Fig. 4}\\
Survival probability for {\it BUND} futures with delivery date:
September  1997.\\
The Mittag-Leffler function (solid line) is compared with the rational	\\
(dash-dotted line) and	the Wright (dashed line)  functions.\\
($\beta =0.95\,,\; \gamma =1/12 $)}}
\end{center}

\vfill\eject

\begin{ack}

F.M and R.G. gratefully acknowledge  for partial support
  the  "Istituto Na\-zio\-na\-le di Alta Ma\-te\-ma\-ti\-ca"
 and the Research Commission of the Free University in Berlin.

\end{ack}

\end{document}